\documentclass{IEEEtran}
\usepackage{amssymb}
\usepackage{amsfonts}
\usepackage{amsmath}
\usepackage{algpseudocode}
\usepackage{algorithm}
\usepackage{caption}
\usepackage{bm}
\usepackage{graphicx,subfigure,cite,multicol,multirow,diagbox,booktabs,array}
\usepackage{stfloats}

\usepackage{color}
\usepackage{float}
\usepackage{subfig}
\usepackage{verbatim}
\usepackage{makecell}
\newenvironment{proof}{\noindent {\textit {\textbf{Proof:}}}}{\hfill$\blacksquare$}
\ifCLASSINFOpdf
\else
\fi
\begin{document}
\title{DNN-based Enhanced DOA Sensing via Massive MIMO Receiver with Switches-based Hybrid Architecture}
\author{Yifan Li, Kang Wei, Linqiong Jia, Jun Zou, Feng Shu,~\emph{Member},~\emph{IEEE}, Yaoliang Song,~\emph{Senior Member},~\emph{IEEE}, and Jiangzhou Wang,~\emph{Fellow},~\emph{IEEE}
}\maketitle	
\begin{abstract}
Switches-based hybrid architecture has attracted much attention, especially in directional-of-arrival (DOA) sensing, due to its ability of significantly reducing the hardware cost by compressing massive multiple-input multiple-output (MIMO) arrays with switching networks.
However, this structure will lead to a degradation in the degrees of freedom (DOF) and accuracy of DOA estimation. To address these two issues, we first propose a switches-based sparse hybrid array (SW-SHA). In this method, we design a dynamic switching network to form a synthesized sparse array, i.e., SW-SHA, that can enlarge the virtual aperture obtained by the difference co-array, thereby significantly enhancing the DOF. Second, in order to improve the DOA estimation accuracy of switches-based hybrid arrays, a deep neural network (DNN)-based method called ASN-DNN is proposed. It includes an antenna selection network (ASN) for optimizing the switch connections based on the criterion of minimizing the Cramér-Rao lower bound (CRLB) under the peak sidelobe level (PSL) constraint and a DNN for DOA estimation. Then by integrating ASN and DNN into an iterative process, the ASN-DNN is obtained. Furthermore, the closed-form expression of CRLB for DOA estimation is derived to evaluate the performance lower bound of switches-based hybrid arrays and provide a benchmark for ASN-DNN. The simulation results show the proposed ASN-DNN can achieve a greater performance than traditional methods, especially in the low signal-to-noise ratio (SNR) regions.
\end{abstract}
\section{Introduction}
Direction of arrival (DOA) estimation is a fundamental and important topic in wireless sensing, and has been applied in various fields like wireless communications, radar, radio astronomy, sonar, navigation, and internet of things (IoT)\cite{krim1996two,tuncer2009classical}. In recent years, with the increasing demands of 5G and 6G for enhancing spectral efficiency, system capacity, and link reliability, massive MIMO has gradually become a key technology\cite{ammar2022user}. At the same time, the application of massive MIMO technology can also provide ultra-high spatial resolution and performance gains for DOA estimation. Therefore, more and more researches on DOA estimation is based on massive MIMO systems\cite{heath2016overview,Cheng2015subspace}.

Although massive MIMO technology can bring significant performance improvements to DOA estimation, as the array aperture expands, the corresponding hardware and software complexity will also increase sharply for the large number of circuit elements, such as radio frequency (RF) chain and analog-to-digital converter (ADC), and large size data matrices to be processed. Therefore, the hybrid analog and digital (HAD) architectures are proposed to decrease the complexity by compressing the large arrays via phase shifter or switch networks\cite{chuang2015high,sohrabi2016hybrid,shu2018low}. Among them, due to the fact that the number of hardware components is much less than other HAD structures, the switches-based hybrid arrays are particularly effective in reducing hardware costs\cite{mendez2016hybrid}.
The work in \cite{zhang2021direction} also proved that switches-based hybrid arrays have a better trade-off between estimation accuracy and power consumption than hybrid arrays with phase shifter network. 

In addition to the positive effect of reducing hardware costs, the application of switches-based hybrid architecture also brings two negative issues. The first is the reduction of array aperture, which is measured by degree of freedom (DOF). Due to the limitation of the number of RF chains, the number of signal sources that the hybrid array can resolve is much fewer than that of a fully digital (FD) array. For example,
As an $M$-elements FD uniform linear array (ULA) is compressed to a $K$-chains hybrid array via a switch network, where $M\gg K$, the DOF of this array for DOA estimation is also compressed from $M-1$ to $K-1$ in theory. 
To enhance the DOF of the arrays, sparse structures are commonly adopted. Common sparse arrays include
minimum redundancy array\cite{moffet1968minimum}, nested array\cite{pal2010nested}, coprime array\cite{vaidyanathan2010sparse}, etc. On the basis of these classical sparse arrays, combined with HAD structures, a series of hybrid sparse arrays have been developed \cite{guo2018doa,rajamaki2020hybrid,koochakzadeh2020compressed}. However, these methods only enhance the DOF to a limited extent. Therefore, this paper considers leveraging the flexibility of switch networks to propose a sparse design based on the switches-based hybrid structure, aiming to significantly increase the virtual aperture of HAD arrays.

The second issue with switches-based hybrid arrays is their relatively weaker DOA estimation performance. In this architecture, each RF chain can only be connected to one antenna, which results in lower spatial accuracy compared to other hybrid structures that utilize phase shifter networks. There are typically two categories of approaches to address this issue. The first category is to perform antenna selection optimization to enhance the spatial resolution of the array configuration. Common optimization strategies include minimizing mean squared error and minimizing the CRLB \cite{gershman1997note,roy2013sparsity,wang2014adaptive}. The second category is to develop higher-performance DOA estimation algorithms, such as \cite{chen2022antenna,zhang2021direction}. Among these, methods based on deep learning (DL) are particularly noteworthy. They have been widely used in DOA estimation due to their powerful learning and reasoning capabilities, as well as their high robustness. DL methods can bring significant performance improvements, especially in low SNR environments, and can also handle the large amount of data brought by massive MIMO systems \cite{liu2018direction,hu2020low,ma2022deep}. Therefore, this paper proposes a high-precision estimator based on DNN, improving the DOA estimation performance of switches-based hybrid arrays from both aspects of optimizing antenna selection and enhancing algorithm performance.

In response to the low DOF and low precision issues in DOA estimation of the switch-based hybrid structure discussed earlier, this paper proposes a novel dynamic sparse structure design method and a DNN-based high-performance DOA estimator.
The main contributions of this paper are summarized as follows:
\begin{enumerate}
	\item To improve the DOF of switches-based hybrid array, a novel sparse array namely SW-SHA is proposed. By leveraging the advantage of switches-based hybrid arrays that can flexibly change array configurations to design a dynamic switching network, and construct a new sparse subarray in each time slot. Then by combining all the subarrays, a synthesized sparse array, i.e., SW-SHA is obtained. It is proven that the difference co-array of SW-SHA is a filled ULA, so the corresponding DOF can be significantly improved. Simulation results also demonstrate that SW-SHA not only has a much higher DOF than existing methods, but also has better DOA estimation performance than them.
	\item To enhance the DOA estimation accuracy of switches-based hybrid array, a DNN-based method called ASN-DNN is proposed. ASN-DNN is consist of two parts. The first part, i.e., ASN, is a neural network designed to select the optimal connection method of the switching network for a specific DOA, based on the criterion of minimizing the CRLB. Additionally, we incorporate the PSL constraint into this criterion to enhance the performance of ASN-DNN under low SNR conditions. The second part, i.e., DNN, is a regression network designed to obtain DOA estimation results. By integrating these two parts and forming an iterative process, then the ASN-DNN method is obtained. Simulation results indicate that ASN-DNN has a significant performance advantage over traditional methods, especially in low SNR regions, and this advantage is robust to changes in angle.
	\item To evaluate the ideal performance of switches-based hybrid array and provide a benchmark for the proposed ASN-DNN method, the closed-form expression of CRLB for DOA estimation is derived in this work. Based on this expression, we formulate an optimization problem aimed at minimizing the CRLB. Upon examining the objective function within this problem, it is discovered that the antenna selection that minimizes the CRLB for a switch-based hybrid array must include the boundary antennas at both sides of the array.
\end{enumerate}

The remainder of this paper is organized as follows. Section \ref{system model} introduces the system model. Section \ref{SW-SHA} describes the design principle of SW-SHA and analyzes the DOF. In Section \ref{section_ASN_DNN}, the DNN-based high-precision DOA estimator ASN-DNN is proposed. Section \ref{crlb analysis} derives the expression of CRLB for switches-based hybrid array. Finally, simulation results and
conclusions are given in Sections \ref{simulation} and \ref{conclusion} respectively.

\emph{\rm{\textbf{Notation}}:}Matrices, vectors, and scalars are denoted by letters of bold upper case, bold lower case, and lower case, respectively. Signs $(\cdot)^T$, $(\cdot)^*$ and $(\cdot)^H$ represent transpose, conjugate and conjugate transpose. $\mathbf{I}$ and $\textbf{0}$ denote the identity matrix and matrix filled with zeros. $\textrm{tr}(\cdot)$ stands for trace operation, $\otimes$ and $\circ$ denotes Kronecker product and Khatri-Rao product. $\left\lfloor \cdot \right \rfloor$ denotes round down.

\section{System Model}\label{system model}
Consider $Q$ far-field narrowband signals impinging on an $M$-elements uniform linear array (ULA). The $q$-th signal is denoted by $s_q(t)e^{j2\pi f_c t}$, where $s_q(t)$ and $f_c$ represent baseband signal and carrier frequency respectively. Then the received signal vector $\mathbf{x}(t)\in \mathbb{C}^{M\times 1}$ is expressed as 
\begin{equation}
		\mathbf{x}(t)=\mathbf{A}(\boldsymbol{\theta})\mathbf{s}(t)+\mathbf{v}(t),
\end{equation} 
where $\mathbf{s}(t)=[s_1(t),\cdots,s_Q(t)]^T$, $\mathbf{v}(t)\sim\mathcal{CN}(\boldsymbol{0},\sigma_v^2\mathbf{I}_M)$ represents the additive white Gaussian noise (AWGN) vector. $\mathbf{A}(\boldsymbol{\theta})=[\mathbf{a}(\theta_1),\cdots,\mathbf{a}(\theta_Q)]\in \mathbb{C}^{M\times Q}$ is the array manifold matrix, and
\begin{equation}
\mathbf{a}(\theta_q)=[e^{j\frac{2\pi}{\lambda} d_0\sin \theta_q},e^{j\frac{2\pi}{\lambda} 2d_0\sin \theta_q},\cdots,e^{j\frac{2\pi}{\lambda}Md_0\sin \theta_q}]^T
\end{equation} 
where $d_0=\lambda/2$ and the ULA index set is given as $\mathbb{P}_{\rm{ULA}}=\{1,2,\cdots,M\}$. 
	
\begin{figure}[tb!]
	\centering
	\includegraphics[width=0.4\textwidth]{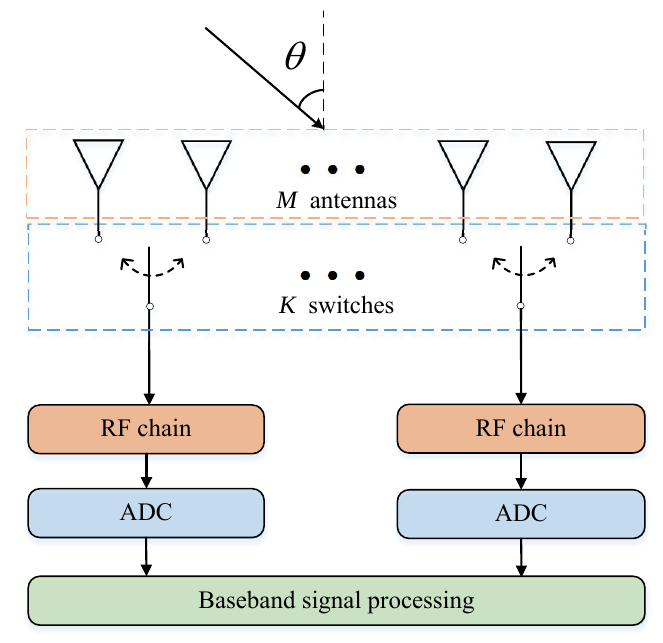}\\
	\caption{Massive MIMO receiver with switches-based hybrid architecture.\label{receive_array}}
\end{figure}

In order to reduce the hardware cost, the switches-based hybrid architecture is adopted in this work. As depicted in Fig.\ref{receive_array}, each antenna connected to a RF chain via a switch, then the $M$-elements ULA is compressed to a $K$-elements array ($K\ll M$) and the output signal $\mathbf{y}(t)\in \mathbb{C}^{K\times 1}$ of the compressed array is given by
\begin{equation}
		\begin{aligned}
		\mathbf{y}(t)=\mathbf{W}^T\mathbf{A}\mathbf{s}(t)+\mathbf{W}^T\mathbf{v}(t)=\tilde{\mathbf{A}}\mathbf{s}(t)+\tilde{\mathbf{v}}(t)
		\end{aligned}
\end{equation}
where $\mathbf{W}=[\mathbf{w}_1,\cdots,\mathbf{w}_K]\in \mathbb{C}^{M\times K}$ is a binary selection matrix, where $\lVert\mathbf{w}_k\lVert_1=1$, and if $k$-th switch is connected to $m$-th antenna we get $\mathbf{W}(m,k)=1$. $\tilde{\mathbf{A}}=\mathbf{W}^T\mathbf{A}=[\tilde{\mathbf{a}}(\theta_1),\cdots,\tilde{\mathbf{a}}(\theta_Q)]\in \mathbb{C}^{K\times Q}$. So $\mathbf{y}(t)$ can be viewed as a signal received by a $K$-elements linear array, and 
\begin{equation}
	\tilde{\mathbf{a}}(\theta_q)=[e^{j\frac{2\pi}{\lambda} \tilde{p}_1d_0\sin \theta_q},e^{j \frac{2\pi}{\lambda} \tilde{p}_2d_0\sin \theta_q},\cdots,e^{j\frac{2\pi}{\lambda} \tilde{p}_Kd_0\sin \theta_q}]^T
\end{equation} 
is the equivalent array manifold vector, and $\tilde{\mathbb{P}}=\{\tilde{p}_1,\tilde{p}_2,\cdots,\tilde{p}_K\}\subset \mathbb{P}$.

Assume that the $Q$ signals follow the stochastic model and are uncorrelated to each other. The noise $\mathbf{v}(t)$ is also independent of signals, so the covariance matrix of $\mathbf{y}(t)$ can be expressed as
\begin{equation}
		\mathbf{R}={\rm{E}}\left[\mathbf{y}(t)\mathbf{y}^H(t)\right]=\tilde{\mathbf{A}}\mathbf{R}_s\tilde{\mathbf{A}}^H+\sigma_v^2\mathbf{I}_K\in\mathbb{C}^{K\times K},\label{covariance_matrix}	
\end{equation}
where $\mathbf{R}_s={\rm{E}}[\mathbf{s}(t)\mathbf{s}^H(t)]=\rm{diag}\{\sigma_1^2,\cdots,\sigma_Q^2\}$ and $\sigma_q^2, q=1,2,\cdots,Q$, denotes the power of $q$-th signal. In practice, $\mathbf{R}$ is usually unavailable for a finite number of snapshots, so we can use the sample covariance matrix
\begin{equation}
	\tilde{\mathbf{R}}=\frac{1}{T}\sum_{t=1}^{T}\mathbf{y}(t)\mathbf{y}^H(t)
\end{equation}  
as an approximation and
when $N\rightarrow \infty$ we get $\tilde{\mathbf{R}}=\mathbf{R}$.

\section{Sparse Design for Switches-based Hybrid Array}\label{SW-SHA}
In order to improve the DOF of hybrid arrays, we design a novel switches-based sparse hybrid array (SW-SHA), and its advantages on DOF are also analyzed in this section.
\subsection{SW-SHA}
\begin{figure}[t!]
	\centering
	\includegraphics[width=0.5\textwidth]{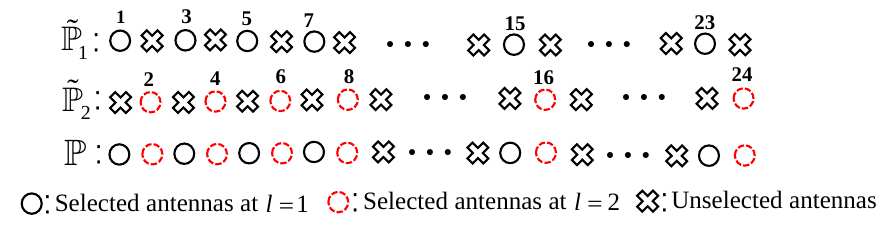}\\
	\caption{The design of SW-SHA with $L=2$ and $K=6$.\label{nested_array}}
\end{figure}

Assume $\mathbf{W}_l$ is a variable selection matrix and it has $L$ different forms, i.e. $l=1,2,\cdots,L$. 
The compressed array generated by $\mathbf{W}_l$ is called subarray $l$ and its output signal is defined as $\mathbf{y}_l(t)=\mathbf{y}(t+(l-1)\tau)$, where 
\begin{equation}
	\begin{aligned}
		\mathbf{y}_l(t)&=\mathbf{W}_l^H\mathbf{A}\mathbf{s}_l(t)+\mathbf{W}_l^H\mathbf{v}(t)\\
		&=\tilde{\mathbf{A}}_l\mathbf{s}(t)e^{j2\pi f_c(l-1)\tau}+\tilde{\mathbf{v}}_l(t),
	\end{aligned}\label{ylt}
\end{equation}
then by multiplying $e^{-j2\pi f_c(l-1)\tau}$ in both sides, (\ref{ylt}) is changed to
\begin{equation}
	\begin{aligned}
	\tilde{\mathbf{y}}_l(t)=\tilde{\mathbf{A}}_l\mathbf{s}(t)+e^{-j2\pi f_c(l-1)\tau}\tilde{\mathbf{v}}_l(t)
	\end{aligned}
\end{equation}
where $\tau< T_s$ denotes the switch delay, $\mathbf{s}_l(t)=\mathbf{s}(t+(l-1)\tau)$ and we get $\mathbf{s}(t+\tau)=\mathbf{s}(t)e^{j2\pi f_c\tau}$ with narrowband assumption \cite{tuncer2009classical} \cite{qin2019doa}.   $\tilde{\mathbf{A}}_l=[\tilde{\mathbf{a}}_l(\theta_1),\cdots,\tilde{\mathbf{a}}_l(\theta_Q)]$, $\tilde{\mathbf{a}}_l(\theta_q)=[e^{j\frac{2\pi}{\lambda} \tilde{p}_{l,1}d_0\sin \theta_q},e^{j \frac{2\pi}{\lambda} \tilde{p}_{l,2}d_0\sin \theta_q},\cdots,e^{j\frac{2\pi}{\lambda} \tilde{p}_{l,K}d_0\sin \theta_q}]^T$, and $\tilde{\mathbb{P}}_l=\{\tilde{p}_{l,1},\tilde{p}_{l,2},\cdots,\tilde{p}_{l,K}\}$ denotes subarray $l$. As shown in Fig.\ref{nested_array}, let all the subarrays in nested form, then $\tilde{\mathbb{P}}_l$ is further expressed by (\ref{p_l}), where $K_1+K_2=K$.
\begin{figure*}[ht]
	\begin{equation}
		\begin{aligned}
			\tilde{\mathbb{P}}_l&=\left\{(k_1-1)L+l~|~1\leq k_1\leq K_1\right\}\cup\left\{(k_2(K_1+1)-1)L+l~|~1\leq k_2\leq K_2\right\}\\
			&=\left\{l,L+l,\cdots,(K_1-1)L+l,K_1L+l,(2K_1+1)L+l\cdots,(K_2(K_1+1)-1)L+l\right\}
		\end{aligned}\label{p_l}
	\end{equation}
\end{figure*}

By combining all the $L$ measurements, we obtain
\begin{equation}
	\begin{aligned}
	\tilde{\mathbf{y}}(t)=\begin{bmatrix}
		\tilde{\mathbf{y}}_1(t)\\
		\tilde{\mathbf{y}}_2(t) \\
		\vdots \\
		\tilde{\mathbf{y}}_L(t) 
	\end{bmatrix}&=\begin{bmatrix}
	\tilde{\mathbf{A}}_1\\
	\tilde{\mathbf{A}}_2 \\
	\vdots \\
	\tilde{\mathbf{A}}_L
\end{bmatrix}\mathbf{s}(t)+\begin{bmatrix}
\tilde{\mathbf{v}}_1(t)\\
e^{-j2\pi f_c\tau}\tilde{\mathbf{v}}_2(t) \\
\vdots \\
e^{-j2\pi f_c(L-1)\tau}\tilde{\mathbf{v}}_L(t) 
\end{bmatrix}\\
&=\tilde{\mathbf{A}}_{\mathbb{P}}\mathbf{s}(t)+\tilde{\mathbf{v}}_{\mathbb{P}}
\end{aligned}
\end{equation}
where $\tilde{\mathbf{y}}(t)\in \mathbb{C}^{LK\times 1}$, $\tilde{\mathbf{A}}_{\mathbb{P}}\in \mathbb{C}^{LK\times Q}$, $\mathbb{P}=\tilde{\mathbb{P}}_1\cup\tilde{\mathbb{P}}_2\cup\cdots\cup\tilde{\mathbb{P}}_L$ represents the augmented array and $|\mathbb{P}|=LK$, $|\cdot|$ denotes the cardinality of set. The covariance matrix of $\tilde{\mathbf{y}}(t)$ is given by
\begin{equation}
	\mathbf{R}_{\mathbb{P}}=\tilde{\mathbf{A}}_{\mathbb{P}}\mathbf{R}_s\tilde{\mathbf{A}}_{\mathbb{P}}^H+\sigma_v^2\mathbf{I}_{LK}.\label{tilde_R}
\end{equation}
As ${\mathbb{P}}$ is a nested-like sparse array, then the difference co-array of ${\mathbb{P}}$ is defined as
\begin{equation}
	\mathbb{D}=\mathbb{D}_s\cup\mathbb{D}_c=\left\{n_1-n_2~\big|~n_1,n_2\in{\mathbb{P}}\right\},\label{difference_coarray}
\end{equation}
where $|\mathbb{D}|=L^2K^2$, $\mathbb{D}_s$ and $\mathbb{D}_c$ denote self-difference set and cross-difference set respectively. Arrange the elements in $\mathbb{D}$ from small to large, and get $\mathbb{D}=\{D_1,D_2,\cdots,D_{L^2K^2}\}$.
By vectorizing (\ref{tilde_R}) and using equation ${\rm{vec}}(\mathbf{A}\mathbf{B}\mathbf{C})=(\mathbf{C}^T\otimes\mathbf{A}){\rm{vec}}(\mathbf{B})$, we get
\begin{equation}
	\begin{aligned}
		\mathbf{r}_{\mathbb{P}}&={\rm{vec}}(\mathbf{R}_{\mathbb{P}})=\left(\tilde{\mathbf{A}}_{\mathbb{P}}^*\circ \tilde{\mathbf{A}}_{\mathbb{P}}\right)\mathbf{r}_s+\sigma^2_v\mathbf{i}_{LK}=\tilde{\mathbf{A}}_{\mathbb{D}}\mathbf{r}_s+\sigma^2_v\mathbf{i}_{LK},
		\label{virtual_signal}
	\end{aligned}
\end{equation}
where $\mathbf{r}_{\mathbb{P}}\in \mathbb{C}^{L^2K^2\times 1}$, $\circ$ represents Katri-Rao product and $\tilde{\mathbf{A}}_{\mathbb{D}}=[\tilde{\mathbf{a}}^*_{\mathbb{P}}(\theta_1)\otimes\tilde{\mathbf{a}}_{\mathbb{P}}(\theta_1),\cdots,\tilde{\mathbf{a}}^*_{\mathbb{P}}(\theta_Q)\otimes\tilde{\mathbf{a}}_{\mathbb{P}}(\theta_Q)]\in \mathbb{C}^{L^2K^2\times Q}$. $\mathbf{r}_s=[\sigma_1^2,\cdots,\sigma_Q^2]^T$ and $\mathbf{i}_K={\rm{vec}}(\mathbf{I}_K)$.

As (\ref{virtual_signal}) can be viewed as a single-snapshot virtual signal, it can be solved by CS-based methods for $Q<L^2K^2$ and the most widely used is on-grid methods. Firstly, given a set of angle grid points $\bar{\boldsymbol{\theta}}=\{\bar{\theta}_1,\cdots,\bar{\theta}_{\bar{Q}}\}$, where $\bar{Q}\gg Q$, $\bar{\theta}_{\bar{q}+1}-\bar{\theta}_{\bar{q}}=\Delta\theta$ and this means the subsequent DOA estimation can only take values from $\bar{\boldsymbol{\theta}}$. Then we get a dictionary matrix $\bar{\mathbf{A}}_{\mathbb{D}}=[\mathbf{a}_{\mathbb{D}}(\bar{\theta}_1),\cdots,\mathbf{a}_{\mathbb{D}}(\bar{\theta}_{\bar{Q}})]^T\in\mathbb{C}^{L^2K^2\times \bar{Q}}$ and the optimization problem is given by
\begin{equation}
	\min_{\bar{\mathbf{r}}_s}~\lVert\bar{\mathbf{r}}_s\rVert_0~~{\rm{s.t.}}~\left\lVert\mathbf{r}_{\mathbb{P}}-	\bar{\mathbf{A}}_{\mathbb{D}}\bar{\mathbf{r}}_s\right\rVert_2\leq \epsilon
	\label{objective_function}
\end{equation}
where $\bar{\mathbf{r}}_s\in \mathbb{C}^{\bar{Q}\times 1}$ is optimization variable and $\lVert\bar{\mathbf{r}}_s\rVert_0=\{\bar{q}:\bar{\mathbf{r}}_s(\bar{q})\neq 0\}$ counts the nonzero entries of $\bar{\mathbf{r}}_s$. $\epsilon$ is a bound related to noise power $\sigma^2_v$. However, (\ref{objective_function}) is NP hard to solve, 
so by introducing LASSO algorithm \cite{tibshirani1996regression}, the objective function (\ref{objective_function}) is transformed to
\begin{equation}
	\min_{\bar{\mathbf{r}}_s}\left[\alpha\lVert\bar{\mathbf{r}}_s\rVert_1+\frac{1}{2}\left\lVert\mathbf{r}_{\mathbb{P}}-	\bar{\mathbf{A}}_{\mathbb{D}}\bar{\mathbf{r}}_s\right\rVert_2\right]
	\label{convex_function}
\end{equation}
where $\alpha$ is a parameter used to make a trade-off between the reconstruction performance and the sparsity of $\bar{\mathbf{r}}_s$. Obviously, the convex optimization problem (\ref{convex_function}) can be directly solved by CVX toolbox. In this work, to improve computation efficiency, we use alternating direction method of multipliers (ADMM) algorithm\cite{boyd2011distributed} as an alternative method.
Then the LASSO problem can be written in the ADMM form
\begin{equation}
	\begin{aligned}
		&\min_{\bar{\mathbf{r}}_s,\mathbf{z}}~f(\bar{\mathbf{r}}_s)+g(\mathbf{z})\\
		&~{\rm{s.t.}}~~\bar{\mathbf{r}}_s=\mathbf{z}
	\end{aligned}
\end{equation}
where $f(\bar{\mathbf{r}}_s)=\frac{1}{2}\left\lVert\mathbf{r}_{\mathbb{P}}-\bar{\mathbf{A}}_{\mathbb{D}}\bar{\mathbf{r}}_s\right\rVert_2$ and $g(\mathbf{z})=\alpha\lVert\mathbf{z}\rVert_1$. According to \cite{boyd2011distributed}, this optimization problem can be solved by following steps:
\begin{subequations}
	\begin{align}
		&\bar{\mathbf{r}}_s^{(i+1)}=\left(\bar{\mathbf{A}}_{\mathbb{D}}^H\bar{\mathbf{A}}_{\mathbb{D}}+\zeta\mathbf{I}_{\bar{Q}}\right)^{-1}\left[\bar{\mathbf{A}}_{\mathbb{D}}^H\mathbf{r}_{\mathbb{P}}+\zeta\left(\mathbf{z}^{(i)}-\mathbf{u}^{(i)}\right)\right],\\
		&\mathbf{z}^{(i+1)}=\mathcal{S}_{\alpha/\zeta}\left(\bar{\mathbf{r}}_s^{(i+1)}+\mathbf{u}^{(i)}\right),\\
		&\mathbf{u}^{(i+1)}=\mathbf{u}^{(i)}+\bar{\mathbf{r}}_s^{(i+1)}-\mathbf{z}^{(i+1)},
	\end{align}
\end{subequations}
where $\zeta>0$, $i$ denotes the number of iterations, and $\mathbf{u}\in\mathbb{R}^{\bar{Q}\times1}$ can be initialized as zero vector. $\mathcal{S}_{\alpha/\zeta}(\cdot)$ is soft threshold operator and is defined as
\begin{equation}
	\mathcal{S}_{\alpha/\zeta}(x)=\begin{cases}
		x-\alpha/\zeta,\quad x>\alpha/\zeta\\
		0, \quad\quad\quad\quad -\alpha/\zeta\leq x\leq \alpha/\zeta\\
		x+\alpha/\zeta, \quad a<-\alpha/\zeta
	\end{cases}.
\end{equation}

\subsection{DOF Analysis}
Since the DOF of a sparse array is related to the architecture of its difference co-array, then we first give the following proposition. 

\noindent{\textit {\textbf{Proposition 1:}}} The difference co-array of augmented array $\mathbb{P}$ is a filled ULA.

\begin{proof}
According to (\ref{p_l}) and (\ref{difference_coarray}), the self-difference set for $\tilde{\mathbb{P}}_l$ is defined as $\mathbb{D}_s(l)=\left\{n_1-n_2~\big|~n_1,n_2\in{\tilde{\mathbb{P}}_l}\right\}$ and select the unique elements from it to form a new set
\begin{equation}
	\mathbb{U}_s(l)=\left\{0, L,2L,\cdots,(K_2(K_1+1)-1)L\right\},
\end{equation}
where $\mathbb{U}_s(l)$ is a symmetric set centered on 0, and the negative part is omitted.
It can be seen that $\mathbb{U}_s(l)$ is not related to $l$, so we get $\mathbb{U}_s(1)=\cdots=\mathbb{U}_s(L)=\mathbb{U}_s$. Then the cross-difference for $\tilde{\mathbb{P}}_{l_1}$ and $\tilde{\mathbb{P}}_{l_2}$ is defined as $\mathbb{D}_c(l_1,l_2)=\left\{n_1-n_2~\big|~n_1\in{\tilde{\mathbb{P}}_{l_1}},n_2\in{\tilde{\mathbb{P}}_{l_2}}\right\}$, and the positive unique elements are selected
\begin{equation}
	\begin{aligned}
	\mathbb{U}_c(l_1,l_2)=&\{l_2-l_1,L+l_2-l_1,\cdots,\\
	&(K_2(K_1+1)-1)L+l_2-l_1\},
	\end{aligned}
\end{equation}
where $1\leq l_1< l_2\leq L$ and $1\leq l_2-l_1\leq L-1$. Therefore, $\mathbb{U}^+=\mathbb{U}_s\cup\mathbb{U}_c=\{0,1,2,\cdots,K_2(K_1+1)L-1\}$ is a filled ULA, i.e. the difference co-array of $\mathbb{P}$ is a filled ULA.
\end{proof} 

On the basis of proposition 1, we select the consecutive unique lags from $\mathbb{D}$ and get the following set
\begin{equation}
	\mathbb{U}=\left\{0,\pm 1,\cdots,\pm (K_2(K_1+1)L-1)\right\},
\end{equation}
and $|\mathbb{U}|=2K_2(K_1+1)L-1$. Then according to \cite{pal2010nested}, we know the DOF of $\mathbb{D}$ is ${\rm{DOF}}=K_2(K_1+1)L-1$. It is evidently that ${\rm{DOF}}$ depend on the value of $L$, so we want to find the maximum $L$ that maximizes it.
Then firstly We arrange the elements in $\mathbb{P}$ from small to large as $\mathbb{P}=\{p_1,p_2,\cdots,p_{LK}\}$ and $p_{LK}=\tilde{p}_{L,K}=K_2(K_1+1)L$. Since the actual aperture of receive array is $M$, then $p_{LK}$ have to satisfy the condition
\begin{equation}
		K_2(K_1+1)L\leq M \Rightarrow L\leq \frac{M}{K_2(K_1+1)},
\end{equation}
and referring to the optimal value of $K_1$, $K_2$ \cite{pal2010nested}, we can get the maximum DOF of the proposed SW-SHA is given in table \ref{DOF},
and DOF of ordinary $K$-chains hybrid array is $K-1$, so SW-SHA has significantly improvement on DOF.

\begin{table}[tb!]
	\centering
	\caption{DOF of the proposed SW-SHA} 
	\begin{tabular}{|c|c|c|}
		\hline
		& $K$ is even & $K$ is odd\\
		\hline
		$K_1$,$K_2$ & $K_1=K_2=K/2$ & \makecell[c]{$K_1=(K-1)/2$\\$K_2=(K+1)/2$} \\
		\hline
		$L_{\rm{max}}$ & $\left\lfloor 4M/(K^2+2K) \right \rfloor$ & $\left\lfloor 4M/(K+1)^2 \right \rfloor$ \\
		\hline
		${\rm{DOF}}$ & $L(K^2+2K)/4-1$ & $L(K+1)^2/4-1$ \\
		\hline
	\end{tabular}\label{DOF}
\end{table}

\section{DNN-based High Resolution DOA Estimator for Switches-based Hybrid Arrays}\label{section_ASN_DNN}
In this section, to improve the DOA estimation of switches-based hybrid array, a DNN-based method which consists of an antenna selection network (ASN) and a DNN estimator is proposed.
\subsection{Antenna Selection Network}
For a switches-based hybrid array with $M$ antennas and $K$ RF chains, there are total $N={M\choose K}$ different selection matrices, i.e.,
\begin{equation}
	\mathcal{W}=\left\{\mathbf{W}_1,\cdots,\mathbf{W}_N\right\}.
\end{equation}
It's obvious that different selection matrices will result in different array beampatterns, and also cause fluctuation of DOA estimation performance. Therefore, in order to achieve high-precision results as far as possible, it's necessary to make an optimization of antenna selection matrix.

To find the optimal antenna selection for a specific DOA $\theta$, the first step is to establish an objective function. Because this section focuses on how to improve the accuracy of DOA estimation, the CRLB for DOA estimation of switches-based hybrid array is utilized here, and its expression is given in (\ref{crlb}). CRLB is a lower bound of estimation error, the lower its value, the higher DOA estimation precision can be potentially achieved with that array configuration. Therefore, the optimal antenna selection matrix of DOA $\theta$ is chosen based on that can make the corresponding CRLB lowest, and the optimization problem is given as
\begin{equation}
	\hat{\mathbf{W}}=\arg\max_{\mathbf{W}\in\mathcal{W}}\left[K\lVert\mathbf{W}^H\mathbf{P}\rVert_{2}^2-\lVert\mathbf{W}^H\mathbf{P}\rVert_{1}^2\right], 
\end{equation}
where $\hat{\mathbf{W}}$ is the optimal selection matrix and $\mathbf{P}=[1,2,\cdots,P]^T$. However, numerous studies have indicated that antenna selection criteria based on minimizing the CRLB can only ensure optimal DOA estimation performance under high signal-to-noise ratio (SNR) conditions \cite{gershman1997note,roy2013sparsity,wang2014adaptive}. When the SNR is low, the array pattern constructed using this method is prone to sidelobe interference. Therefore, to enhance the accuracy of DOA estimation under low SNR conditions, refer to the design in \cite{wang2014adaptive}, we consider adding a constraint on the peak sidelobe level (PSL). Firstly, the definition of PSL is 
\begin{equation}
	{\rm{PSL}}=\frac{V_s}{V_m}
\end{equation}
where 
\begin{equation}
	V_s=\big|\mathbf{a}^H(\theta_s)\mathbf{W}\mathbf{W}^H\mathbf{a}(\theta_m)\big|,
\end{equation}
\begin{equation}
	V_m=\big|\mathbf{a}^H(\theta_m)\mathbf{W}\mathbf{W}^H\mathbf{a}(\theta_m)\big|,
\end{equation}
$\theta_m$ and $\theta_s$ denotes the directions of mainlobe and sidelobe respectively. So the optimization problem is transformed to
\begin{equation}
	\begin{aligned}
		&\max_{\mathbf{W}\in\mathcal{W}} K\lVert\mathbf{W}^H\mathbf{P}\rVert_{2}^2-\lVert\mathbf{W}^H\mathbf{P}\rVert_{1}^2\\
		& ~~{\rm{s.t.}} \quad {\rm{PSL}}\leq \delta, 
	\end{aligned}
\end{equation}
where $\delta\in [0,1]$ is the desired normalised sidelobe power level with respect to the mainlobe. Then the search procedure of optimal antenna selection for $\theta$ is: 1) set $\theta$ as mainlobe direction and evaluate the PSLs of antenna selection matrices in $\mathcal{W}$, 2) hold the antenna selection matrices that meet the PSL constraint and the matrix can minimize CRLB is optimal.

\begin{figure}[tb!]
	\centering
	\includegraphics[width=0.5\textwidth]{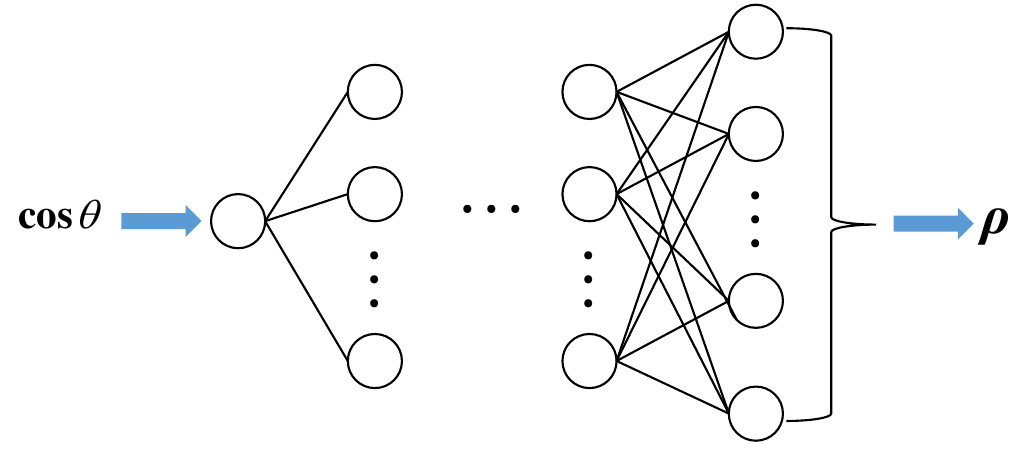}\\
	\caption{Network structure of ASN.\label{ASN}}
\end{figure}

Furthermore, in a massive MIMO system, the value of $N$ might be very large, and the time cost of traversing the set $\mathcal{W}$ entirely is too high for a real-time DOA estimation task. So in order to enhance the timeliness, we propose a neural network-based ASN to solve the antenna selection problem, which can reduce the time cost during the optimization process by offline training. As shown in Fig. \ref{ASN}, the proposed ASN has $H$ hidden layers, and each layer contains $g_h$, $1\leq h\leq H$ neurons, We define $g_0=1$ and $g_{H+1}=M$ as the number of neurons contained in input layer and output layer respectively. Based on the objective function of (\ref{crb_optimization}), we choose $\cos\theta$ as the input of ASN and the output of ASN is a $M$-elements selection vector $\boldsymbol{\rho}\in \{0,1\}^M$ which is defined as
\begin{equation}
	\boldsymbol{\rho}=\sum_{k=1}^K \mathbf{w}_k,
\end{equation}
where $\mathbf{w}_k$ is the $k$-th column of antenna selection matrix $\mathbf{W}$. Then the output of the $h$th hidden layer can be obtained as
\begin{equation}
	\mathbf{x}_h={\rm{ReLu}}\left(\mathbf{U}_h\mathbf{x}_{h-1}+\mathbf{b}_h\right),
\end{equation}
where $\mathbf{U}_h\in\mathbb{R}^{g_{h-1}\times g_h}$ and $\mathbf{b}_h\in\mathbb{R}^{g_{h}\times 1}$ denote weight matrix and bias vector, ${\rm{ReLu}}(\cdot)$ means this layer adopts rectified linear unit (ReLu) as activation function. For the output layer, in consideration of the output vector $\boldsymbol{\rho}$'s form (binary and multi-label), we adopt Sigmoid function as activation function and the final output of ASN is given by
\begin{equation}
	\mathbf{x}_{H+1}={\rm{Sigmoid}}\left(\mathbf{U}_{H+1}\mathbf{x}_{H}+\mathbf{b}_{H+1}\right).
\end{equation}
Then by combining all the layers, the computation process of ASN can be expressed by
\begin{equation}
	\begin{aligned}
		\hat{\boldsymbol{\rho}}&=f_{\rm{ASN}}\left(\cos\theta,\mathbf{u}_{\rm{ASN}}\right)\\
		&={\rm{Sigmoid}}\left({\rm{ReLu}}\left(\cdots {\rm{ReLu}}\left(\mathbf{U}_1\cos\theta+\mathbf{b}_1\right)\right)\right),\label{ASN_process}
	\end{aligned}
\end{equation} 
where $\mathbf{u}_{\rm{ASN}}$ is a vector that contains all the weights and biases of ASN.

In the training stage, we sample the angular range $[\theta_{\rm{min}},\theta_{\rm{max}})$ at intervals of $\Delta\theta$ to obtain $N_{\theta}$ points, where 
\begin{equation}
	N_{\theta}=\frac{\theta_{\rm{max}}-\theta_{\rm{min}}}{\Delta\theta},
\end{equation}
and construct an angular set $\Theta=\{\theta_1,\cdots,\theta_{N_{\theta}}\}$. Then calculate the optimal antenna selection vector corresponding to each angle in $\Theta$, and obtain a training dataset
\begin{equation}
	\mathcal{T}_{\rm{ASN}}=\left\{\left\langle\cos\theta_i,\check{\boldsymbol{\rho}}_i\right\rangle|1\leq i\leq N_{\theta}\right\},
\end{equation}
where 
\begin{equation}
	\check{\boldsymbol{\rho}}_i=\sum_{k=1}^K \check{\mathbf{w}}_{i,k},
\end{equation}
where $\check{\mathbf{w}}_{i,k}$ is the $k$-th column of optimal antenna selection matrix $\check{\mathbf{W}}_i$ for $\theta_i$. During the training procedure, the model will export a probability distribution-form vector $\hat{\boldsymbol{\rho}}_i$, we want the difference between it and label vector as small as possible, so we introduce binary cross-entropy (BCE) as loss function, which is defined as following
\begin{equation}
	\begin{aligned}			L_{\rm{ASN}}=-\frac{1}{N_{\theta}}\sum_{i=1}^{N_{\theta}}\sum_{m=1}^{M}\bigg[\check{\rho}_{i,m}\log(\hat{\rho}_{i,m})+\\
		(1-\check{\rho}_{i,m})\log(1-\hat{\rho}_{i,m})\bigg],
	\end{aligned}
\end{equation}
where $\check{\rho}_{i,m}$ and $\hat{\rho}_{i,m}$ denote the $m$th element of $\check{\boldsymbol{\rho}}_i$ and $\hat{\boldsymbol{\rho}}_i$. Then by minimizing $L_{\rm{ASN}}$ the optimal parameters of ASN can be obtained. In the test stage, import an arbitrary angle into ASN, the corresponding indexes of $K$ biggest values in the output vector are the selected antenna indexes, so the antenna selection problem can be resolved timely.

\subsection{DNN-based DOA Estimator}
\begin{figure}[tb!]
	\centering
	\includegraphics[width=0.5\textwidth]{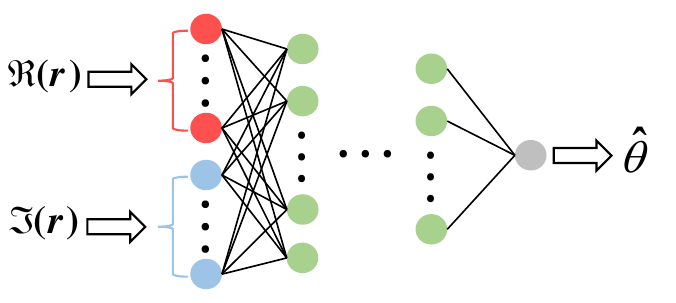}\\
	\caption{Network structure of DNN estimator.\label{DNN}}
\end{figure}

After antenna selection, the next step is DOA estimation. Similarly, to improve the DOA estimation performance, we also propose a DNN-based method as depicted in Fig. \ref{DNN}. For extracting useful information from the received signals, we choose the elements of the received signal covariance matrix $\mathbf{R}$ as the input for the DNN. Since the lower left elements of $\mathbf{R}$ are conjugate replicas of the upper right ones, so we vectorize $\mathbf{R}$ and abandon the redundant elements, then get the following vector 
\begin{equation}
	\bar{\mathbf{r}}=\left[R_{1,1},\cdots,R_{1,K},\cdots,R_{K,K}\right]^T\in\mathbb{C}^{\frac{K(K+1)}{2}\times1},
\end{equation}
where $R_{i,j}$ denotes the element at $i$th row and $j$th column of $\mathbf{R}$. However, the input for a neural network must be real, $\bar{\mathbf{r}}$ can't be directly used. Then we extract the real part and image part of $\bar{\mathbf{r}}$ respectively, and construct a new vector as
\begin{equation}
	\mathbf{r}=\left[\Re(\bar{\mathbf{r}});\Im(\bar{\mathbf{r}})\right]\in\mathbb{R}^{{K(K+1)}\times1}.
\end{equation}
So the input layer of DNN contains $K(K+1)$ neurons. To avoid the limitation of DOA estimation accuracy by grid constraints, a regression strategy is employed to design the DNN here, meaning that the DNN exports the true angle rather than a category. Consequently, the output layer of this DNN contains only a single neurons. Based on the angular set $\Theta$ generated in ASN section, the training dataset for DNN can be given as
\begin{equation}
	\mathcal{T}_{\rm{DNN}}=\left\{\left\langle\mathbf{r}_i,\theta_i\right\rangle|1\leq i\leq N_{\theta}\right\}.
\end{equation}

Ignoring the assumptions made in the previous section, the number of hidden layers in the DNN is set to $H$ here. Then the computation process of DNN is expressed by
\begin{equation}
	\begin{aligned}
		\hat{\theta}&=f_{\rm{DNN}}\left(\mathbf{r},\mathbf{u}_{\rm{DNN}}\right)\\
		&=f_{H+1}\left(f_H\left(\cdots f_1\left(\mathbf{U}_1\mathbf{r}+\mathbf{b}_1\right)\right)\right),
	\end{aligned}
\end{equation}
where $\mathbf{u}_{\rm{DNN}}$ is the parameter vector of DNN and $f_h(\cdot)$ represents activation function. We choose ReLU as the activation function for the hidden layers, while the output layer uses a linear activation function due to the regression task. Subsequently, the mean squared error (MSE) is adopted as the loss function to measure the discrepancy between the DNN's output and the labels during the training process, which is defined as:
\begin{equation}
	L_{\rm{DNN}}=\frac{1}{N_{\theta}}\sum_{i=1}^{N_{\theta}}\left(\hat{\theta}_i-\theta_i\right)^2.
\end{equation}
So by minimizing the above loss function, we can obtain the optimal parameters of DNN estimator.

\begin{figure}[tb!]
	\centering
	\includegraphics[width=0.5\textwidth]{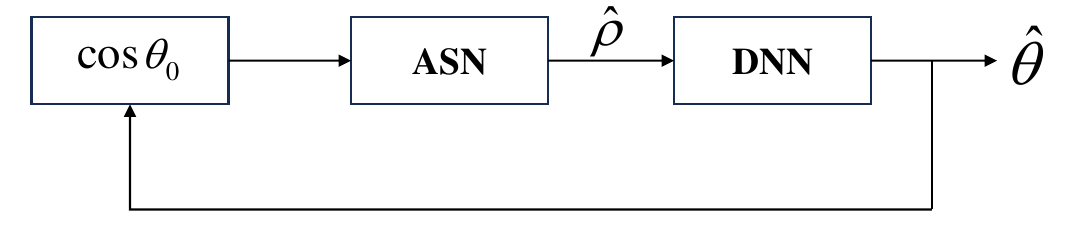}\\
	\caption{Schematic diagram for ASN-DNN.\label{ASN-DNN}}
\end{figure}

\subsection{ASN-DNN}
In the first two sections, we proposed antenna selection and DOA estimation methods based on DNN, and now we will introduce how to combine these two parts to obtain the optimal DOA estimation results. As depicted in Fig. \ref{ASN-DNN}, ASN initially takes a coarsely estimated angle as input, then outputs the result of the antenna selection. Following this outcome, the array structure is adjusted, and feature vectors are extracted from the received signals and fed into DNN. The angle output by the DNN is subsequently sent back to the ASN for the next round of optimization. This cycle continues until specific termination conditions are met.

Because it is an iterative process, we need to select a suitable initial value for the proposed ASN-DNN method to ensure that the estimation accuracy and convergence speed of the method are both maintained at a high level. Therefore, the method for calculating the initial value must have the characteristics of high precision and low complexity. Here, we use the Minimum Variance Distortionless Response (MVDR) algorithm to address this, and its expression is as follows:
\begin{equation}
	P(\theta)=\frac{1}{\mathbf{a}^H(\theta)\tilde{\mathbf{R}}^{-1}\mathbf{a}(\theta)},\label{mvdr}
\end{equation}
by searching the peak of $P(\theta)$ we can get a DOA estimation result. The specific procedure of ASN-DNN is presented in algorithm \ref{ASN-DNN_algorithm}.

\begin{algorithm}[htb!]
	\caption{ASN-DNN}\label{ASN-DNN_algorithm}
	\begin{algorithmic}
		\Require $\boldsymbol{\rho}_{0}$: initial antenna selection vector;
		\State 1: Adjust array configuration based on $\boldsymbol{\rho}_{0}$, and compute the initial angle $\theta_0$ via (\ref{mvdr});
		\State 2: Given $\hat{\theta}_0=\theta_0$ and $j=1$, then
		\State 3: \textbf{while} {$|\hat{\theta}_j-\hat{\theta}_{j-1}|\leq \varepsilon$} \textbf{do}
		\State 4: \quad  Input $\cos\theta_{j-1}$ to pre-trained ASN, obtain the optimal antenna selection vector $\hat{\boldsymbol{\rho}}_j$;
		\State 5: \quad Adjust configuration based on $\boldsymbol{\rho}_{j}$, and construct feature vector $\mathbf{r}_j$;
		\State 6: \quad Input $\mathbf{r}_j$ to pre-trained DNN, obtain DOA estimation result $\hat{\theta}_j$;
		\State 7: \quad $j=j+1$;
		\State 8: \textbf{end while}
		\State 9: Assume the above iteration ends at $j=J$, the final DOA estimation result is $\hat{\theta}=(\hat{\theta}_J+\hat{\theta}_{J-1})/2$;
		\Ensure $\hat{\theta}$.
	\end{algorithmic}
\end{algorithm}

\section{CRLB Analysis}\label{crlb analysis}
In this section, we derive the closed-form expression of CRLB for the DOA estimation via switches-based hybrid array and make some analysis on it. CRLB is a lower bound on the variance of any unbiased estimator, so it can provide a benchmark for evaluating the DOA estimation performance of the proposed ASN-DNN. Furthermore, the closed-form expression of CRLB can be helpful for understanding the impact of array architectures on the DOA estimation performance. Then the expression of CRLB is given in the following theorem.

\textit{Theorem 1:} The CRLB expression for DOA estimation via switches-based hybrid architecture is given by
\begin{equation}
	\begin{aligned}
		{\rm{CRLB}}_{\boldsymbol{\theta}}
=\frac{\sigma_v^2}{2T}\left\{{\rm{Re}}\left[\left(\dot{\tilde{\mathbf{A}}}^H\boldsymbol{\Pi}_{\tilde{\mathbf{A}}}^{\perp}\dot{\tilde{\mathbf{A}}}\right)\odot\left(\mathbf{R}_s\tilde{\mathbf{A}}^H\mathbf{R}^{-1}\tilde{\mathbf{A}}\mathbf{R}_s\right)\right]\right\}^{-1},\label{CRB}
	\end{aligned}
\end{equation}
where
\begin{subequations}
	\begin{align}
		&\boldsymbol{\Pi}_{\boldsymbol{\Sigma}}^{\perp}=\mathbf{I}_{K^2}-\mathbf{F}_{\boldsymbol{\Sigma}}\left(\mathbf{F}_{\boldsymbol{\Sigma}}^H\mathbf{F}_{\boldsymbol{\Sigma}}\right)^{-1}\mathbf{F}_{\boldsymbol{\Sigma}}^H,	\\
		&\boldsymbol{\Pi}_{\tilde{\mathbf{A}}}^{\perp}=\mathbf{I}_{K}-\tilde{\mathbf{A}}\left(\tilde{\mathbf{A}}^H\tilde{\mathbf{A}}\right)^{-1}\tilde{\mathbf{A}}^H,\\
		&\dot{\tilde{\mathbf{A}}}=\frac{\partial \tilde{\mathbf{A}}}{\partial \boldsymbol{\theta}^T}=\left[\frac{\partial \tilde{\mathbf{a}}(\theta_1)}{\partial \theta_1},\cdots,\frac{\partial \tilde{\mathbf{a}}(\theta_Q)}{\partial \theta_Q}\right].
	\end{align}
\end{subequations}
\textit{Proof:} See Appendix A. $\hfill\blacksquare$

In the single source case, we can get the simplified CRLB expression and obtain useful insights for the DOA estimation with switch-based hybrid architecture. As $Q=1$, then the covariance matrix of $\mathbf{y}$ is given by
\begin{equation}
	\mathbf{R}=\sigma_s^2\tilde{\mathbf{a}}(\theta)\tilde{\mathbf{a}}^H(\theta)+\sigma_v^2\mathbf{I}_K,\label{covariance_matrix}	
\end{equation}
and the FIM of $\theta$ is simplified to
\begin{equation}
	\begin{aligned}
		F_{\theta}&=\frac{2(\sigma_s^2)^2}{\sigma_v^2}\dot{\tilde{\mathbf{a}}}^H\boldsymbol{\Pi}_{\tilde{\mathbf{a}}}^{\perp}\dot{\tilde{\mathbf{a}}}\tilde{\mathbf{a}}^H\mathbf{R}^{-1}\tilde{\mathbf{a}}\label{c}
	\end{aligned}
\end{equation}
where $\tilde{\mathbf{a}}=\tilde{\mathbf{a}}(\theta)$ and
\begin{equation}
	\begin{aligned}
		\dot{\tilde{\mathbf{a}}}=\frac{\partial \tilde{\mathbf{a}}(\theta)}{\partial \theta}=j\frac{2\pi d_0}{\lambda}\cos\theta\mathbf{D}\tilde{\mathbf{a}}(\theta),
	\end{aligned}	
\end{equation}
where $\mathbf{D}={\rm{diag}}\{\tilde{p}_1,\cdots, \tilde{p}_K\}$. Then we can get
\begin{equation}
	\begin{aligned}
		\dot{\tilde{\mathbf{a}}}^H\boldsymbol{\Pi}_{\tilde{\mathbf{a}}}^{\perp}\dot{\tilde{\mathbf{a}}}&=\frac{4\pi^2d_0^2\cos^2\theta}{\lambda^2 }\tilde{\mathbf{a}}^H\mathbf{D}\left(\mathbf{I}_{K}-\tilde{\mathbf{a}}\left(\tilde{\mathbf{a}}^H\tilde{\mathbf{a}}\right)^{-1}\tilde{\mathbf{a}}^H\right)\mathbf{D}\tilde{\mathbf{a}}\\
		&=\frac{4\pi^2d_0^2\cos^2\theta}{\lambda^2 }\left[\tilde{\mathbf{a}}^H\mathbf{D}\mathbf{D}\tilde{\mathbf{a}}-\frac{1}{K}\tilde{\mathbf{a}}^H\mathbf{D}\tilde{\mathbf{a}}\tilde{\mathbf{a}}^H\mathbf{D}\tilde{\mathbf{a}}\right]\\
		&=\frac{4\pi^2d_0^2\cos^2\theta}{K\lambda^2 }\left[K\sum_{k=1}^K\tilde{p}_k^2-\left(\sum_{k=1}^K\tilde{p}_k\right)^2\right], \label{a}
	\end{aligned}
\end{equation}
where $\tilde{\mathbf{a}}^H\tilde{\mathbf{a}}=K$. And by using Sherman-Woodbury identity \cite{tuncer2009classical}, $\tilde{\mathbf{a}}^H\mathbf{R}^{-1}\tilde{\mathbf{a}}$ can be further solved as
\begin{equation}
	\begin{aligned}
		\tilde{\mathbf{a}}^H\mathbf{R}^{-1}\tilde{\mathbf{a}}&=\frac{1}{\sigma_v^2}\tilde{\mathbf{a}}^H\left(\mathbf{I}_K-\frac{\gamma\tilde{\mathbf{a}}\tilde{\mathbf{a}}^H}{1+\gamma K}\right)\tilde{\mathbf{a}}\\
		&=\frac{K}{\sigma_v^2(1+K\gamma)}.\label{b}
	\end{aligned}
\end{equation}
where $\gamma=\sigma_s^2/\sigma_v^2$.

Therefore, by substituting (\ref{a}) and (\ref{b}) into (\ref{c}), the expression of $F_{\theta}$ can be obtained as
\begin{equation}
	\begin{aligned}
		F_{\theta}=\frac{8\pi^2d_0^2\gamma^2\cos^2\theta}{\lambda(1+K\gamma) }\left[K\sum_{k=1}^K\tilde{p}_k^2-\left(\sum_{k=1}^K\tilde{p}_k\right)^2\right],
	\end{aligned}
\end{equation}
then the expression of CRLB is given as
\begin{equation}
	\begin{aligned}
		{\rm{CRLB}}_{\theta}&=\frac{1}{T}F_{\theta}^{-1}\\
		&=\frac{1}{T\beta}\left[K\sum_{k=1}^K\tilde{p}_k^2-\left(\sum_{k=1}^K\tilde{p}_k\right)^2\right]^{-1},
	\end{aligned}
\end{equation}
where
\begin{equation}
	\beta=\frac{8\pi^2d_0^2\gamma^2\cos^2\theta}{\lambda(1+K\gamma) }.
\end{equation}

Let $\mathbf{P}=[1,2,\cdots,M]^T$ and $\boldsymbol{\Pi}={\rm{diag}}\{\mathbf{P}\}$, where $\boldsymbol{\Pi}(m,m)=\mathbf{P}(m)$, $1\leq m\leq M$. Then the expression of ${\rm{CRLB}}_{\theta}$ can be transformed to
\begin{equation}
	\begin{aligned}
	{\rm{CRLB}}_{\theta}&=\frac{1}{T\beta}\left[K\lVert\mathbf{W}^H\mathbf{P}\rVert_{2}^2-\lVert\mathbf{W}^H\mathbf{P}\rVert_{1}^2\right]^{-1}\\
	&=\frac{1}{T\beta}\left[K\lVert\boldsymbol{\Pi}\boldsymbol{\rho}\rVert_{2}^2-\lVert\boldsymbol{\Pi}\boldsymbol{\rho}\rVert_{1}^2\right]^{-1}.\label{crlb}
	\end{aligned}
\end{equation}

Obviously, from the above expression, we can see when the signal direction $\theta$ is fixed, parameter $\beta$ is constant. So, the minimum ${\rm{CRLB}}_{\theta}$ can be achieved via solving the following optimization problem
\begin{equation}
	\begin{aligned}
		\max_{\boldsymbol{\rho}}~&{K\big\lVert\boldsymbol{\Pi}\boldsymbol{\rho}\big\rVert_{2}^2-\big\lVert\boldsymbol{\Pi}\boldsymbol{\rho}\big\rVert_{1}^2}\\
		~{\rm{s.t.}}~~&\big\lVert\boldsymbol{\rho}\big\rVert_{1}=K,~~\rho_m\in\{0,1\}\label{crb_optimization}.
	\end{aligned}
\end{equation}
By observing the objective function of (\ref{crb_optimization}) and combining the experiment results in \cite{gershman1997note}, we can get a remark as follow:

\textit{Remark 1:} The optimal antenna selection that minimizes the CRLB must include the boundary antennas at the two sides of the array.

%
%
%

\section{Simulation Results}\label{simulation}
In this section, some simulation results will be presented to evaluate the performance of the proposed SW-SHA and ASN-DNN methods in this paper.
\subsection{Performance Evaluation of SW-SHA}
In this section, we evaluate the DOA estimation performance of SW-SHA via numerical simulations. And we also introduce nested array\cite{pal2010nested}, coprime array\cite{vaidyanathan2010sparse} and compressed sparse array (CSA)\cite{guo2018doa} as benchmarks for SW-SHA, where nested array and Coprime array are classical sparse arrays, CSA is a hybrid array with sparse architecture. If not specifically stated, all the sources are assumed to have same power and part of the simulation parameters are set as: $M=128$, $K=8$, $L=6$, $T=600$, $\alpha=0.25$ and $\Delta \theta=1^{\circ}$. Table \ref{DOF_simu} gives the DOF of these considered array structures under this parameter condition. Obviously, under the same number of RF chains, the DOF of SW-SHA is much higher than that of other array structures, more than six times that of the nested array, and close to four times that of CSA. With the flexibility in array configuration, switches-based hybrid arrays have achieved a huge advantage in DOF. Therefore, compared to the fully digital arrays with high hardware cost and other hybrid structures that cannot easily change their configurations, switches-based hybrid arrays have the potential to resolve the DOA estimation problems from more signal sources at a lower cost.

\begin{figure}[tb!]
	\centering  
	\subfigure[SW-SHA]{
		\includegraphics[width=0.41\textwidth]{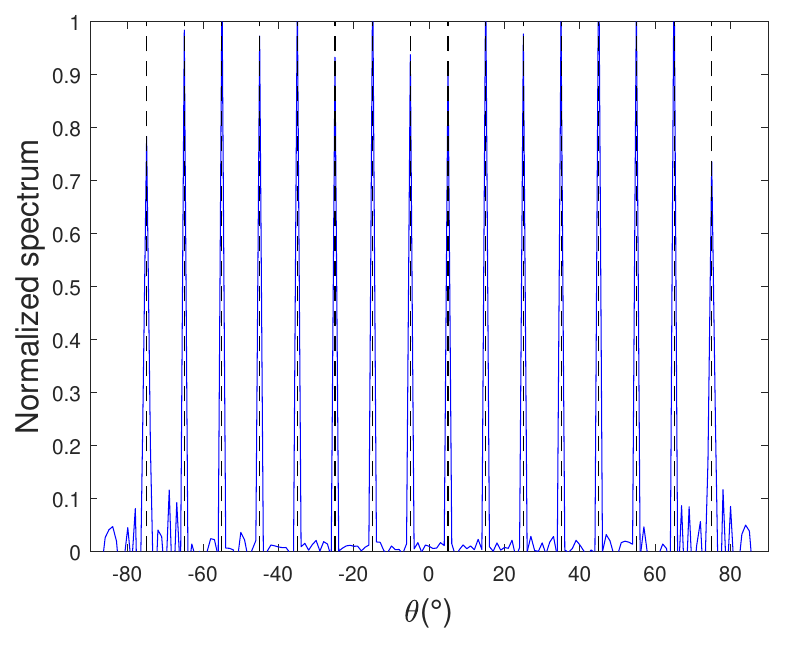}}\\
	\subfigure[Nested array]{
		\includegraphics[width=0.39\textwidth]{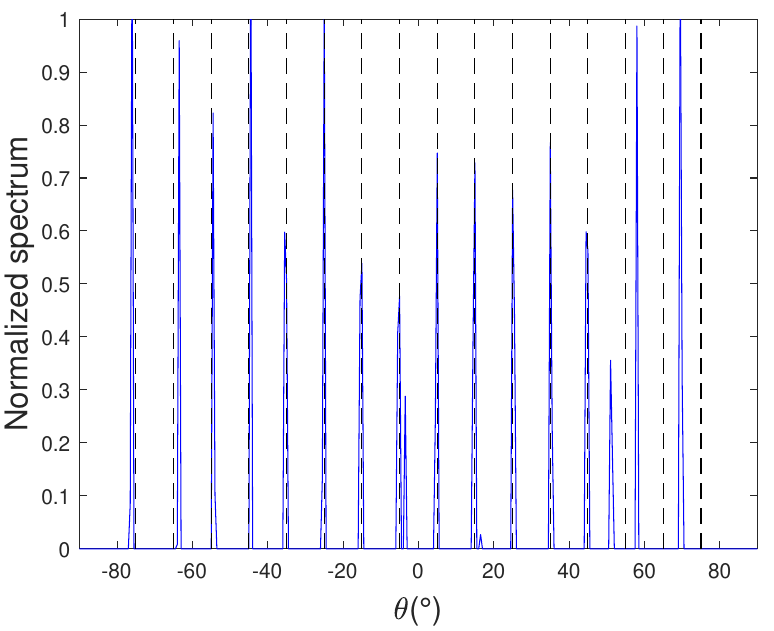}}\\
	\subfigure[CSA]{
		\includegraphics[width=0.41\textwidth]{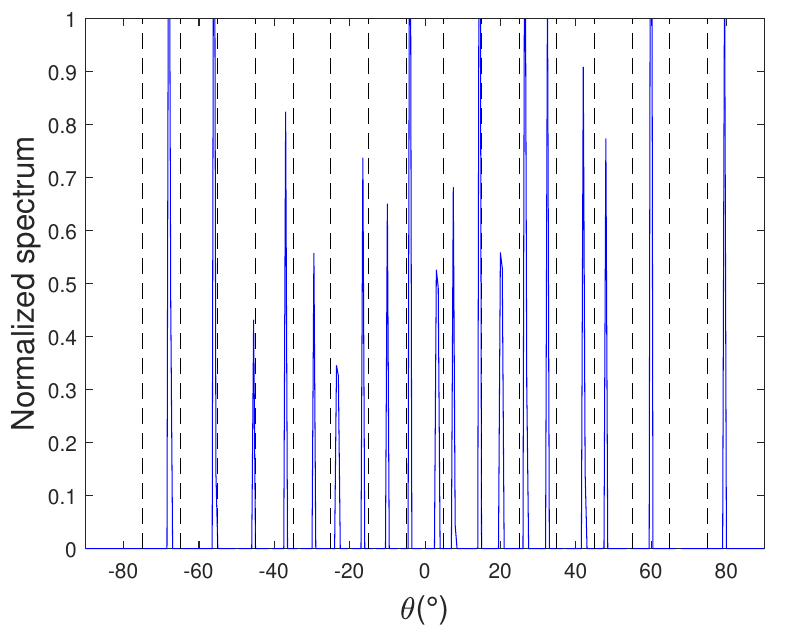}}\\
	\caption{The spatial spectrums of different array architectures.}\label{spatial spectrum}
\end{figure}

Fig.\ref{spatial spectrum} shows the spatial spectrums of these considered arrays with $Q=16$, the black dashed lines indicate the preset directions of the signal sources, and the peaks of the blue curve represent the estimated directions. The spatial spectrum of an array can reflect its ability to resolve the number of signal sources and the accuracy of DOA estimation. It can be seen from Fig. \ref{spatial spectrum} (a) that the spatial spectrum peaks of SW-SHA completely coincide with the preset directions of the signal sources. For the nested array and CSA, although the number of signal sources is less than their DOF, their spatial spectra cannot accurately predict the preset angles. Therefore, SW-SHA not only theoretically increases the value of DOF but also ensures the quality of DOA estimation in practical applications.

\begin{table}[tb!]
	\centering
	\caption{DOF of different array architectures} 
	\begin{tabular}{|c|c|c|c|c|c|}
		\hline
		& SW-SHA & Nested array & Coprime array & CSA & ULA\\
		\hline
		DOF & 119 & 19 & 15 & 31 & 7 \\
		\hline
	\end{tabular}\label{DOF_simu}
\end{table}

\begin{figure}[tb!]
	\centering
	\includegraphics[width=0.45\textwidth]{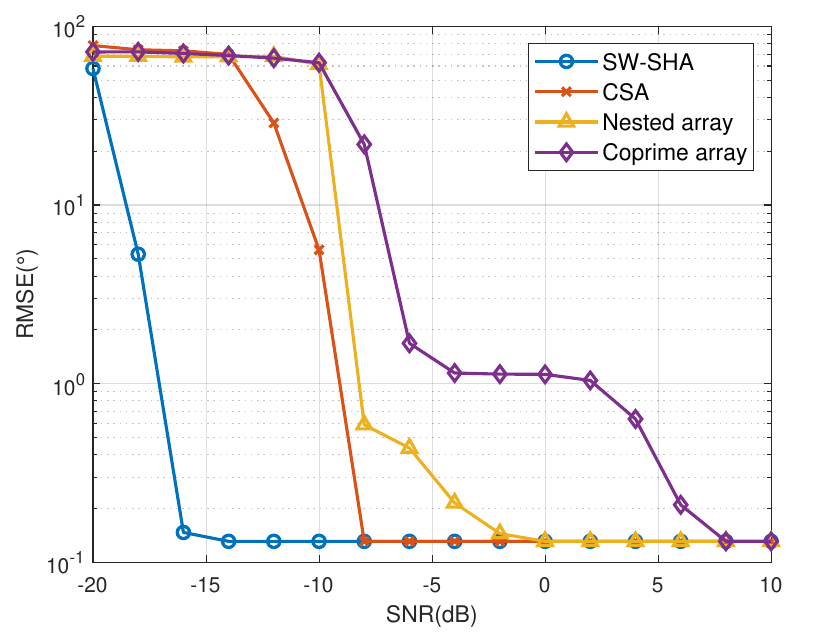}\\
	\caption{RMSE versus SNR.\label{rmse_snr}}
\end{figure}

To further validate the DOA estimation performance of the proposed SW-SHA, Fig.\ref{rmse_snr} presents the curves of RMSE versus SNR with $Q=1$ and $\theta=-67.131^{\circ}$. From this figure we can find the DOA estimation accuracy of SW-SHA is much higher than CSA, nested array and coprime array, especially when $\rm{SNR}<-8dB$, and the largest difference is more than $50^{\circ}$. Combining the Fig. \ref{spatial spectrum} and Fig. \ref{rmse_snr}, it can be seen that SW-SHA outperforms other sparse arrays in both scenarios with large number of signal sources and a single signal source. Therefore, the significance of proposing SW-SHA lies not only in enhancing the DOF of switch-based hybrid arrays, but also in improving the DOA estimation performance of sparse arrays.

\subsection{Performance Evaluation of ASN-DNN}
In this subsection, we evaluate the DOA estimation performance of the proposed ASN-DNN method via numerical simulations. In these simulations, the proposed ASN-DNN is compared to a $K$-antennas ULA with Root-MUSIC estimator and the DNN without antenna selection optimization. Furthermore, the CRLB of switches-based hybrid architecture derived in this work is also introduced as a baseline for the three methods. These simulations were all completed on a desktop computer with Windows 10, AMD Ryzen 7 5800X CPU and NVIDIA GeForce RTX 4070Ti GPU. The software environment setup comprises Python 3.9 and Tensorflow 2.6.0. In the training stage, the angular range is set as $\Theta=[-60^{\circ},60^{\circ}]$ and the angular sample interval is $\Delta \theta=1^{\circ}$. The considered SNR range is $[-15{\rm{dB}},10{\rm{dB}}]$ and the interval is 5dB. The other parameters are same as the set in the simulations of SW-SHA. The DOA estimation performance of the considered methods are evaluated by the root mean-squared error (RMSE) and all results are averaged over 5000 Monte-Carlo simulations, which can be expressed by
\begin{equation}
	{\rm{RMSE}}=\sqrt{\frac{\sum_{i=1}^{5000}(\hat{\theta}_i-\theta)^2}{5000}}.
\end{equation}

\begin{figure}[tb!]
	\centering
	\includegraphics[width=0.45\textwidth]{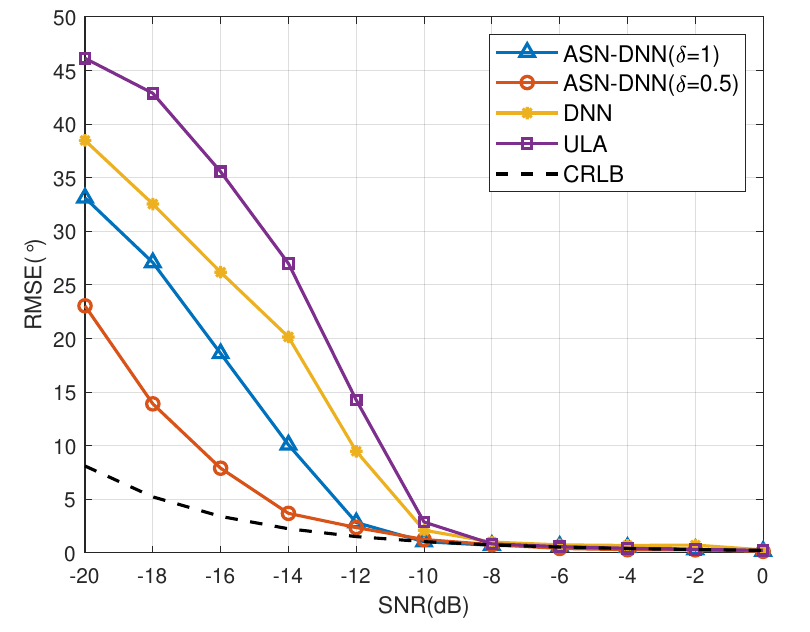}\\
	\caption{RMSE versus SNR.\label{ASN-DNN_rmse_snr}}
\end{figure}

Fig. \ref{ASN-DNN_rmse_snr} shows the relationship between RMSE and SNR when $T=100$ and $\theta=30^{\circ}$. In this simulation, the proposed ASN-DNN is trained with $\delta=1$ and $\delta=0.5$ respectively, where $\delta=1$ means the PSL constraint doesn't work and the antenna selection only relies on minimizing CRLB. The antenna selection results for $\delta=1$ and $\delta=0.5$ are $\mathbb{P}_{\delta=1}=\{1,2,3,4,125,126,127,128\}$ and $\mathbb{P}_{\delta=0.5}=\{1,2,4,8,119,124,127,128\}$. The antenna index of ULA is $\mathbb{P}_{\rm{ULA}}=\{1,2,3,4,5,6,7,8\}$. For the pure DNN, it doesn't have a step of antenna selection optimization, so its antenna index can be generated randomly. Here to compare the performance between the proposed DNN and traditional methods like Root-MUSIC, the antenna index for DNN is set same as ULA, i.e., $\mathbb{P}_{\rm{DNN}}=\mathbb{P}_{\rm{ULA}}$. From Fig. \ref{ASN-DNN_rmse_snr} we can see the proposed ASN-DNN has significant performance advantages at low SNR regions especially when ${\rm{SNR}}\leq -10{\rm{dB}}$ compared to DNN and ULA, so it's obvious that ASN plays a crucial role in enhancing the DOA estimation performance of the switches-based hybrid structure. By comparing the performance of ASN-DNN with different $\delta$, it can be observed that when $\rm{SNR}<-12dB$, the estimation performance of the ASN-DNN with $\delta=0.5$ is much better than that of the ASN-DNN $\delta=1$, where the maximum estimation error difference exceeding $10^{\circ}$. This phenomenon demonstrates that selecting an antenna connection method with a lower sidelobe level can effectively enhance the DOA estimation performance of switches-based hybrid arrays within the low SNR range, which aligns with our expectation when adding the PSL constraint.

\begin{figure}[tb!]
	\centering
	\includegraphics[width=0.455\textwidth]{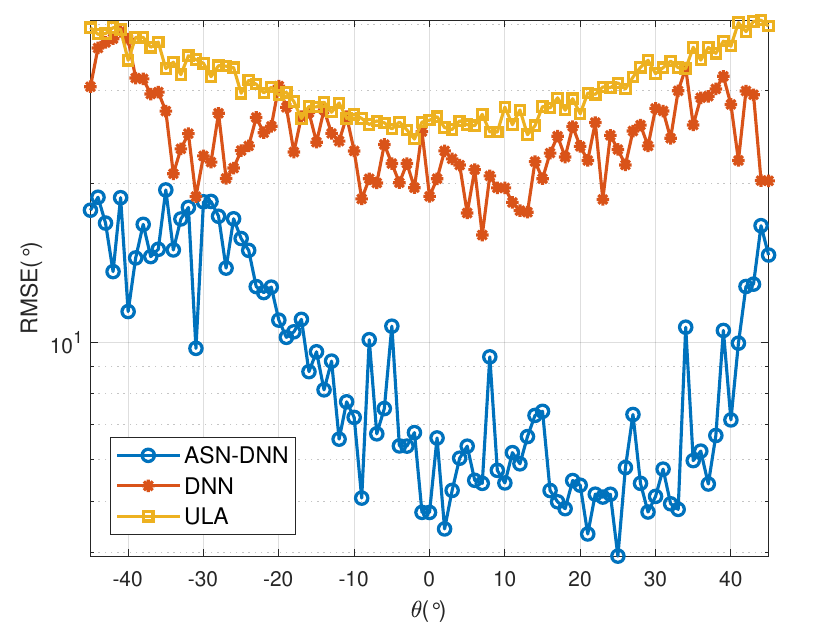}\\
	\caption{RMSE versus $\theta$.\label{ASN-DNN_rmse_theta}}
\end{figure}

\begin{figure}[tb!]
	\centering
	\includegraphics[width=0.45\textwidth]{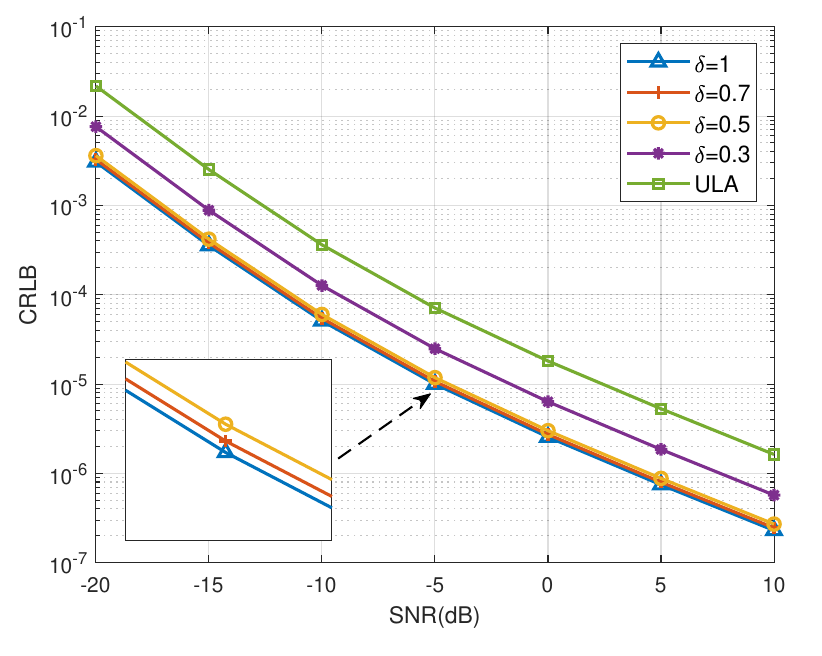}\\
	\caption{CRLBs of switches-based hybrid arrays with optimal antenna selection under different $\delta$.\label{crlb_delta}}
\end{figure}

To observe whether the method proposed in this paper performs consistently at different angles, Fig. \ref{ASN-DNN_rmse_theta} plots the curve of RMSE varying with $\theta$ when $\rm{SNR} = -15 dB$. The considered angular range in this simulation is $[-45^{\circ},45^{\circ}]$, it can be seen that throughout the entire angular range, the DOA estimation error of ASN-DNN is lower than that of DNN and ULA, and this advantage is particularly evident in the range from $-10^{\circ}$ to $40^{\circ}$. Therefore, ASN-DNN has high robustness and can achieve high-precision DOA estimation over a wide angular range.

Finally, we also provide the curves of the CRLBs of the optimal array configurations generated under different $\delta$ varying with SNR in Fig. \ref{crlb_delta}, to verify the impact of the PSL constraint on the theoretical accuracy of ASN-DNN. Since the PSL constraint does not take effect when $\delta = 1$, the corresponding optimal array configuration has the lowest CRLB among all configurations, which can also be seen from the figure. Fig. \ref{crlb_delta} shows that the lower $\delta$, the higher the CRLB of the corresponding optimal array configuration. This means that reducing $\delta$ may potentially lead to an increase in DOA estimation error, especially in the high SNR region. However, we can also observe that when $0.5\leq\delta<1$, the corresponding CRLB differs very little from that when $\delta = 1$. But when $\delta<0.5$ (for example, $\delta = 0.3$), this difference becomes significant. Therefore, we can control $\delta$ to be above 0.5, so as not to have too much impact on the theoretical performance of DOA estimation.

\section{Conclusion}\label{conclusion}
In this work, we first proposed the SW-SHA to improve the DOF of DOA estimation for switches-based hybrid arrays. Leveraging the flexibility of switches and sparse structures, a dynamic switching network is designed to generate a sparse subarray in each time slot. By combining all subarrays, the SW-SHA is constructed. The virtual aperture obtained by the difference co-array of SW-SHA can reach the maximum, thus significantly enhancing the DOF. Compared to traditional sparse arrays, SW-SHA not only makes a greater improvement in DOF but also offers a higher DOA estimation accuracy. Subsequently, ASN-DNN is proposed to enhance the DOA estimation accuracy of switches-based hybrid arrays. The main aim of ASN is to optimize antenna selection based on the criterion of minimizing the CRLB under the PSL constraint, while the DNN serves as a regression network for DOA estimation. Then by integrating ASN and DNN into an alternating iteration procedure, the ASN-DNN method is established. Furthermore, the expression of CRLB for DOA estimation via switches-based hybrid arrays is also derived as a benchmark. Simulation results demonstrate the proposed ASN-DNN may make a dramatic performance improvements over existing methods in the low SNR regions and is robust to the change of signal directions.
\appendices
\renewcommand{\appendixname}{Appendix}
\section{Derivation of CRLB}
Given a random vector $\mathbf{y}$ with probability density function (PDF) $p(\mathbf{y};\boldsymbol{\alpha})$, where $\boldsymbol{\alpha}$ is a parameter vector, its fisher information matrix (FIM) is defined as
\begin{equation}
	\mathbf{F}=-{\rm{E}}\left\{\frac{\partial^2 \ln p(\mathbf{y};\boldsymbol{\alpha})}{\partial\boldsymbol{\alpha}\partial\boldsymbol{\alpha}^T}\right\},
\end{equation}
as the observation vector $\mathbf{y}$ in this work follows zero-mean complex Gaussian distribution with covariance matrix $\mathbf{R}$ and the elements in $\mathbf{F}$ can be expressed as
\begin{equation}
	\mathbf{F}_{i,j}={\rm{tr}}\left\{\mathbf{R}^{-1}\frac{\partial \mathbf{R}}{\partial \alpha_i}\mathbf{R}^{-1}\frac{\partial \mathbf{R}}{\partial \alpha_j}\right\},\quad 1\leq i,j\leq 2Q+1,\label{FIM_element}
\end{equation}
where $\boldsymbol{\alpha}=[\boldsymbol{\theta}^T,\boldsymbol{\Sigma}^T]$ is a $(2Q+1)\times 1$ parameter vector, $\boldsymbol{\theta}=[\theta_1,\cdots,\theta_Q]^T$ and $\boldsymbol{\Sigma}=[\sigma_1^2,\cdots,\sigma_Q^2,\sigma_v^2]^T=[\boldsymbol{\Sigma}_s^T,\sigma_v^2]$. According to the parameter vector, we can get the following partitioned FIM
\begin{equation}
	\mathbf{F}=\begin{bmatrix}
		\mathbf{F}_{\boldsymbol{\theta}\boldsymbol{\theta}} & \mathbf{F}_{\boldsymbol{\theta}\boldsymbol{\Sigma}}\\
		\mathbf{F}_{\boldsymbol{\Sigma}\boldsymbol{\theta}} & \mathbf{F}_{\boldsymbol{\Sigma}\boldsymbol{\Sigma}}
	\end{bmatrix}.
\end{equation}

Based on the following properties
\begin{equation}
	\begin{aligned}
		&\left(\mathbf{A}\otimes\mathbf{B}\right)^{H}=\mathbf{A}^{H}\otimes\mathbf{B}^{H},\\
		&\left(\mathbf{A}\otimes\mathbf{B}\right)^{-1}=\mathbf{A}^{-1}\otimes\mathbf{B}^{-1},\\
		&{\rm{tr}}\left(\mathbf{A}\mathbf{B}\mathbf{C}\mathbf{D}\right)=\left({\rm{vec}}\left(\mathbf{B}^H\right)\right)^H\left(\mathbf{A}^T\otimes\mathbf{C}\right){\rm{vec}}\left(\mathbf{D}\right),\\
	\end{aligned}
\end{equation}
and $\mathbf{R}^H=\mathbf{R}$, equation (\ref{FIM_element}) can be transformed to
\begin{equation}
	\begin{aligned}
		&\mathbf{F}_{i,j}=\left({\rm{vec}}\left(\frac{\partial \mathbf{R}}{\partial \alpha_i}\right)\right)^H\left(\mathbf{R}^{-T}\otimes\mathbf{R}^{-1}\right){\rm{vec}}\left(\frac{\partial \mathbf{R}}{\partial \alpha_j}\right)\\
		&=\left(\frac{\partial \mathbf{r}}{\partial \alpha_i}\right)^H\left(\mathbf{R}^{T}\otimes\mathbf{R}\right)^{-1}\frac{\partial \mathbf{r}}{\partial \alpha_j}\\
		&=\left[\left(\mathbf{R}^{T}\otimes\mathbf{R}\right)^{-1/2}\frac{\partial \mathbf{r}}{\partial \alpha_i}\right]^H\left[\left(\mathbf{R}^{T}\otimes\mathbf{R}\right)^{-1/2}\frac{\partial \mathbf{r}}{\partial \alpha_j}\right],
	\end{aligned}
\end{equation}
where $\mathbf{r}={\rm{vec}}(\mathbf{R})$.
Therefore, the partitioned FIM can be further expressed as 
\begin{equation}
	\begin{aligned}
		\mathbf{F}=\begin{bmatrix}
			\mathbf{F}_{\boldsymbol{\theta}}^H\\
			\mathbf{F}_{\boldsymbol{\Sigma}}^H
		\end{bmatrix}
		\begin{bmatrix}
			\mathbf{F}_{\boldsymbol{\theta}} & \mathbf{F}_{\boldsymbol{\Sigma}}
		\end{bmatrix}=\begin{bmatrix}
			\mathbf{F}_{\boldsymbol{\theta}}^H\mathbf{F}_{\boldsymbol{\theta}} & \mathbf{F}_{\boldsymbol{\theta}}^H\mathbf{F}_{\boldsymbol{\Sigma}}\\
			\mathbf{F}_{\boldsymbol{\Sigma}}^H\mathbf{F}_{\boldsymbol{\theta}} & \mathbf{F}_{\boldsymbol{\Sigma}}^H\mathbf{F}_{\boldsymbol{\Sigma}}
		\end{bmatrix}
	\end{aligned}
\end{equation}
where
\begin{equation}
	\begin{aligned}
		\mathbf{F}_{\boldsymbol{\theta}}&=\left(\mathbf{R}^{T}\otimes\mathbf{R}\right)^{-1/2}\frac{\partial \mathbf{r}}{\partial \boldsymbol{\theta}^T},\\
		\mathbf{F}_{\boldsymbol{\Sigma}}&=\left(\mathbf{R}^{T}\otimes\mathbf{R}\right)^{-1/2}\frac{\partial \mathbf{r}}{\partial \boldsymbol{\Sigma}^T}\\
		&=\left(\mathbf{R}^{T}\otimes\mathbf{R}\right)^{-1/2}\left[\frac{\partial \mathbf{r}}{\partial \boldsymbol{\Sigma}_s^T}\bigg{|}\frac{\partial \mathbf{r}}{\partial \sigma_v^2}\right].\\
	\end{aligned}
\end{equation}
CRLB is defined as the inverse of FIM, and we are only interested in the estimation accuracy of $\boldsymbol{\theta}$, so according to derivation in \cite{stoica2001stochastic}, we can get
\begin{equation}
	\begin{aligned}
		&{\rm{CRLB}}_{\boldsymbol{\theta}}
		=\frac{1}{T}\left(\mathbf{F}_{\boldsymbol{\theta}}^H\boldsymbol{\Pi}_{\boldsymbol{\Sigma}}^{\perp}\mathbf{F}_{\boldsymbol{\theta}}\right)^{-1}\\
		&=\frac{\sigma_v^2}{2N}\left\{{\rm{Re}}\left[\left(\dot{\tilde{\mathbf{A}}}^H\boldsymbol{\Pi}_{\tilde{\mathbf{A}}}^{\perp}\dot{\tilde{\mathbf{A}}}\right)\odot\left(\mathbf{R}_s\tilde{\mathbf{A}}^H\mathbf{R}^{-1}\tilde{\mathbf{A}}\mathbf{R}_s\right)\right]\right\}^{-1}
	\end{aligned}
\end{equation}
where
\begin{subequations}
	\begin{align}
		&\boldsymbol{\Pi}_{\boldsymbol{\Sigma}}^{\perp}=\mathbf{I}_{K^2}-\mathbf{F}_{\boldsymbol{\Sigma}}\left(\mathbf{F}_{\boldsymbol{\Sigma}}^H\mathbf{F}_{\boldsymbol{\Sigma}}\right)^{-1}\mathbf{F}_{\boldsymbol{\Sigma}}^H,	\\
		&\boldsymbol{\Pi}_{\tilde{\mathbf{A}}}^{\perp}=\mathbf{I}_{K}-\tilde{\mathbf{A}}\left(\tilde{\mathbf{A}}^H\tilde{\mathbf{A}}\right)^{-1}\tilde{\mathbf{A}}^H,\\
		&\dot{\tilde{\mathbf{A}}}=\frac{\partial \tilde{\mathbf{A}}}{\partial \boldsymbol{\theta}^T}=\left[\frac{\partial \tilde{\mathbf{a}}(\theta_1)}{\partial \theta_1},\cdots,\frac{\partial \tilde{\mathbf{a}}(\theta_Q)}{\partial \theta_Q}\right].
	\end{align}
\end{subequations}

\bibliographystyle{IEEEtran}
\bibliography{sparse_subarray}
\end{document}